# COVID-19 Evolves in Human Hosts

Yanni Li*
SCST, Xidian University, China
yannili@mail.xidian.edu.cn

Bing Liu*
WICT, Peking University, China.
dcsliub@pku.edu.cn

Zhi Wang
SCST, Xidian University, China
zhiwang@stu.xidian.edu.cn

Jiangtao Cui
SCST, Xidian University, China
cuijt@xidian.edu.cn

Kaicheng Yao
SCST, Xidian University, China
kcyao@stu.xidian.edu.cn

Pengfan Lv
SCST, Xidian University, China
pengflv@stu.xidian.edu.cn

Yulong Shen
SCST, Xidian University, China
ylshen@mail.xidian.edu.cn

Yueshen Xu
SCST, Xidian University, China
ysxu@xidian.edu.cn

Yuanfang Guan
DCMB & MM, UMich, USA
yuanfang.guan.1.0@gmail.com

Xiaoke Ma
SCST, Xidian University, China
xkma@xidian.edu.cn

## ABSTRACT

Today, we are all threatened by an unprecedented pandemic: COVID-19. How different is it from other coronaviruses? Will it be attenuated or become more virulent? Which animals may be its original host? In this study, we collected and analyzed nearly thirty thousand publicly available complete genome sequences for COVID-19 virus from 79 different countries, the previously known flu-causing coronaviruses (HCov-229E, HCov-OC43, HCov-NL63 and HCov-HKU1) and the lethal, pathogenic viruses, SARS, MERS, Victoria, Lassa, Yamagata, Ebola, and Dengue. We found strong similarities between the current circulating COVID-19 and SARS and MERS, as well as COVID-19 in rhinolophines and pangolins. On the contrary, COVID-19 shares little similarity with the flu-causing coronaviruses and the other known viruses. Strikingly, we observed that the divergence of COVID-19 strains isolated from human hosts has steadily increased from December 2019 to May 2020, suggesting COVID-19 is actively evolving in human hosts. In this paper, we first propose a novel *MLCS* algorithm *HA-MLCS*[1] for the big sequence data (with sequence length over $10^3$) analysis, which can calculate the common model for COVID-19 complete genome sequences to provide important information for vaccine and antibody development. Geographic and time-course analysis of the evolution trees of the human COVID-19 reveals possible evolutional paths among strains from 79 countries. This finding has important implications for the management of COVID-19 and the development of vaccines and medications.

---

*Both authors contributed equally to this research.

[1]The source code of *HA-MLCS* is available at: https://github.com/HA-MLCS/HA-MLCS

## 1 INTRODUCTION

Since its first report in December 2019, the severe infectious pneumonia caused by the new COVID-19 virus has spread widely from the Wuhan City, across China, and to 188 countries. On March 11, 2020, the WHO announced COVID-19 outbreak a pandemic, the first of its kind since the 2009 Swine Flu. Internationally, as of May 29, 2020, the outbreak of COVID-19 has resulted in more than 5,851,494 cases and 361,270 deaths[2]. COVID-19 is currently the biggest health, economical and survival threat to the entire human race. We are in urgent need to understand this virus, find treatment and develop vaccines to combat it.

One challenge in developing effective antibodies and vaccines for COVID-19 is that we do not yet understand this virus. How far away is it from other coronaviruses? Has it undergone any changes since its first discovery? These questions are critical for us to find cures and design effective vaccines and medications, and critical to manage this virus. The study of COVID-19 began only recently [1–6]. So far, pioneering studies related to the virus have been limited to a few COVID-19 complete genome sequences (COVID-19 sequences/strains/viruses for short) and a few related viruses [7]. One study used six COVID-19 sequences from patients in Wuhan and compared them with those of SARS and MERS [8]. Another two studies used nine and five sequences respectively, and found that COVID-19 is similar to SARS [9, 10]. Recent work [11] studied the emergence of genomic diversity and recurrent mutations

---

[2]COVID-19 Dashboard by the Center for Systems Science and Engineering (CSSE) at the Johns Hopkins University (JHU)

in COVID-19 by using 7666 public genome assemblies. These pioneering efforts laid the foundation for our work. In this paper, we collected nearly thirty thousand complete genome sequences, covering 29,305 genomes isolated from COVID-19 in human hosts from 79 countries, 21 genomes from animals and the environment (outside the human bodies), 101 genomes from the four previously known flu-causing coronavirus, and 61 genomes from the seven potentially lethal pathogenic viruses. This collection allows us to analyze the evolution and diversity of COVID-19 in depth. Note that, in this paper all computations/analyses are done using only the collected COVID-19 complete genome sequences.

In this paper, we report strong shared similarity between the currently circulating COVID-19 and the SARS virus, as well as strong shared similarities with COVID-19 in rhinolophines (especially with two strains) and in pangolins. On the contrary, COVID-19 shares a moderate sequence similarity to the four flu-causing coronaviruses, despite reported similar symptoms. Strikingly, we observed the divergence of COVID-19 strains isolated from human hosts steadily increased from December 2019 to May 2020, suggesting COVID-19 is now actively evolving in human hosts. This may potentially explain the differences in the death rate in different areas, as the virus might have evolved into strains of different lethality. Importantly, in this paper we first proposed a novel *MLCS* algorithm for the big sequences analysis, which can calculate the common model (common subsequences) for COVID-19 sequences and provide important information for future studies of vaccine and antibody design. Evolutionary analysis of the human COVID-19 from 79 countries reveals the following important discoveries: *As early as Dec. 2019, the COVID-19 virus was widespread in many countries and regions,* and it is particularly worth noting that the entire genome sequences of the top 15 countries with the most severe epidemics, except Russia and Spain, almost do not reside in the first generation on the evolution tree from 79 countries' sequences, which is of great significance to the traceability of COVID-19. Moreover, the other findings by big sequences analysis in this paper may also provide important information to the understanding and the management of COVID-19 and to the development of vaccines and medications for the virus in the near future.

The rest of this paper is organized as follows. Section 2 discusses our proposed novel *MLCS* (<u>M</u>ultiple <u>L</u>ongest <u>C</u>ommon <u>S</u>ubsequence) algorithm *HA-MLCS* for COVID-19 big sequence data similarity analysis. The big data analysis results for COVID-19 strains are reported in Section 3. Section 4 concludes the paper.

## 2 A MLCS ALGORITHM HA-MLCS

### 2.1 Preliminaries

**MLCS Problem.** We define a subsequence of a given sequence over a finite alphabet $\Sigma$ as a sequence obtained by deleting zero or more (not necessarily consecutive) characters from the given sequence. Let $X = x_1 x_2 ... x_n$ and $Y = y_1 y_2 ... y_m$ be two sequences with lengths $n$ and $m$, respectively, over a finite alphabet $\Sigma$, i.e., $x_i, y_j \in \Sigma$. The goal of the <u>L</u>ongest <u>C</u>ommon <u>S</u>ubsequence (*LCS*) mining problem is to find all longest common subsequences of $X$ and $Y$. Similarly, the goal of the *MLCS* problem is to find all longest common subsequences from $d$ ($d \geqslant 3$) sequences of equal length $n$ or different lengths. *LCS* is a special case of *MLCS*.

The *MLCS* problem is a classical NP-hard problem [12], which is related to the identification of sequences similarity and to the common model extraction between sequences. It has many important applications in bioinformatics, computational genomics, pattern recognition, etc. Based on the adopted method, existing *MLCS* algorithms can be classified into two categories: *dynamic programming based* and *dominant-point based* exact or approximate algorithms.

**(1) Dynamic Programming Algorithms.** Given two sequences $X = x_1 x_2 ... x_n$ and $Y = y_1 y_2 ... y_m$ with lengths $n$ and $m$, respectively, over a finite alphabet $\Sigma$ with $X[i] = x_i, Y[j] = y_j, x_i, y_j \in \Sigma$, $1 \leqslant i \leqslant n$ and $1 \leqslant j \leqslant m$, a dynamic programming algorithm iteratively constructs a $(n+1) \cdot (m+1)$ *score matrix* $L$, where $L[i, j]$ is the length of an *LCS* between two prefixes $X'$ and $Y'$ of $X$ and $Y$.

Once the score matrix $L$ is calculated, all the *LCS*s can be obtained by tracing back from the end element $L[n, m]$ to the starting element $L[0, 0]$. Both the time and space complexities of this algorithm are $O(mn)$. Given $d$ ($d \geqslant 3$) sequences with equal or unequal lengths, the matrix $L$ can be extended to $d$ dimensions for the *MLCS* problem, in which the element $L[i_1, i_2, ..., i_d]$ can be calculated in a similar way. Both the time and space complexity is $O(n^d)$ [13].

**(2) Dominant-point Based Algorithms.** The dominant-point based algorithms are motivated by the observation that most of the cells in the score matrix $L$ of the input sequences are useless and do not need to be computed. Only a very small subset of the cells, called dominants (see Def.1 in Sec. 2.3), should be computed and stored. A dominant-point based *MLCS* algorithm consists of two steps [14, 16]: 1) constructing a directed acyclic graph, called *MLCS-DAG* (<u>D</u>irected <u>A</u>cyclic <u>G</u>raph), which consists of all *MLCS*s of input sequences; 2) computing all of the *MLCS*s of the sequences based on the *MLCS-DAG*.

Although many *MLCS* algorithms [13, 16, 17] have shown that the dominant-point based *MLCS* algorithms are much faster than the classical dynamic programming based algorithms, theoretical analysis and some statistical experiments[18, 19] reveal that the current mainstream ***dominant-point based MLCS algorithms are hard to apply to big sequence data*** due to their *MLCS-DAG* with a massive number of redundant points, as well as memory and computation combinatorial explosion for large-scale and/or big sequence data, namely big sequences.

### 2.2 Related Work

Considering the space-time cost, approximate *MLCS* algorithms are usually designed for mining *MLCS*s of big sequences. As we aim to propose a high precision and efficient approximation *MLCS* algorithm in this paper, we only review existing representative approximation *MLCS* algorithms.

Existing approximate *MLCS* algorithms can be divided into two categories: with or without a guaranteed *performance ratio*, the ratio of *MLCS* length (i.e., |*MLCS*|) of an approximate solution to that of the optimal one.

Algorithms such as *LR*, *ExpA*, and *BNMAS* belong to the first category [17]. They all provide a guaranteed performance ratio of $1/|\Sigma|$, where $|\Sigma|$ is the size of the sequence's alphabet $\Sigma$. Although interesting theoretically, they are not very useful in practice because the performance ratio of $1/|\Sigma|$ is too small, e.g., 1/4 for DNA sequences ($|\Sigma| = 4$). Algorithms in the second category usually use heuristic or probabilistic search techniques to achieve a good performance. For example, Shyu and Tsai [20] used ant colony optimization to find approximate solutions. Wang et al. [21] proposed a heuristic greedy search algorithm *MLCS-APP*, and *Pro-MLCS* [17] adopted an iterative best first search strategy to progressively output better and better solutions. Yang et al. [22] presented two space-efficient approximate *MLCS* algorithms, *SA-MLCS* and *SLA-MLCS* with an iterative beam widening search strategy to reduce the space usage during the iterative calculating process. Experiments show that *SA-MLCS* and *SLA-MLCS* can solve an order of magnitude larger size instances than the state-of-the-art approximate algorithm *Pro-MLCS*. Although this second class of algorithms claims that optimal solutions can be found, the quality of the solutions is difficult to evaluate as there are no exact baseline algorithms for comparison.

From the literature review of existing representative approximate *MLCS* algorithms, we make the following observations: 1) These algorithms' precision is too low to meet the practical need; 2) these algorithms are hard to apply to big sequences due to the weakness of their underlying dominant-point based methods; 3) despite great efforts, no approximate *MLCS* algorithm can tackle the big sequence data *MLCS* mining efficiently and effectively. The proposed *HA-MLCS* algorithm aims to achieve this.

## 2.3 A Novel Approximate MLCS Algorithm

**Definition 1**: For a sequence set $T = \{S_1, S_2, ..., S_i, ..., S_d\}$ over a finite alphabet $\Sigma$, and $|S_i| = n$,[3] let $S_i[p_i]$ ($S_i[p_i] \in \Sigma$) be the $p_i$-th character in $S_i$. The point $p = (p_1, p_2, ..., p_d)$ is called a ***matched point*** of $T$, if and only if $S_1[p_1] = S_2[p_2] = ... = S_i[p_i] = ... = S_d[p_d] = c$ ($p_i \in \{1, 2, ..., n\}$, $c \in \Sigma$).

**Definition 2**: For two matched points, $p = (p_1, p_2, ..., p_d)$ and $q = (q_1, q_2, ..., q_d)$ of $T$, if $\forall i, p_i < q_i$, we say that $p$ strongly dominates $q$, denoted by $p \prec q$, where $p$ is referred to as a *dominating point* (*dominant* for short) and $q$ as a ***dominated point or successor*** of $p$. Further, if there is no matched point $r = (r_1, r_2, ..., r_d)$ for $T$ such that $p \prec r \prec q$, we say that $q$ is ***an immediate successor*** of $p$ and $p$ is ***an immediate predecessor*** of $q$.

**Definition 3**: A dominant point $p = (p_1, p_2, ..., p_d)$ is called the *k-th dominant* (the ***k-level point*** for short), if in the *score matrix*, $L[p_1, p_2, ..., p_d] = k$. The set of all $k$-th dominants is denoted as $D^k$, and the set of all dominants of $T$ is denoted as $D$.

In what follows, we'll go into detail on the main procedures of our novel approximate *MLCS* algorithm and its underlying theory.

**Constructing Successor Tables (ST).** To obtain all the immediate successors of a dominate $p$ from sequences $T$ in $O(1)$ time, we design a new data structure, called successor tables *ST* of $T$. The construction and operation of *ST* are detailed in Appendix A.

**Constructing optimized MLCS-DAG.** To overcome the weaknesses of the *MLCS-DAG* of the existing dominant-point based algorithm, we construct an optimized *MLCS-DAG*, called *MLCS-ODAG*, with a minimum number of non-critical points (not contributing to *MLCSs* of sequences set $T$). To this end, we construct its *MLCS-ODAG* with the following procedure:

1) Two dummy $d$-dimensional points $(0, 0, ..., 0)$ (the source point) and $(\infty, \infty, ..., \infty)$ (the sink point) are first introduced into the *MLCS-ODAG* for $d$-dimensional sequences, with all the other points in *MLCS-ODAG* being the successors of $(0, 0, ..., 0)$ and the predecessors of $(\infty, \infty, ..., \infty)$. Initially, let $k = 0$ and $D^k = \{(0, 0, ..., 0)\}$.

2) If $D^k = \emptyset$, goto 6); otherwise, for each point $p$ in $D^k$, calculate all its immediate successors by the successor tables of $T$ and add a directed edge to each of their successors from $p$. If point $p$ has no successor, a directed edge from $p$ to the sink point is added. All of the successors of points from $D^k$ constitute an initial $(k + 1)th$ level point set, denoted as $D^{k+1}_{init}$.

3) To eliminate many redundant points (repeated and non-critical points) possibly residing in $D^{k+1}_{init}$, a retention strategy is employed, that is, only those best points (possible key points for short) that are most likely to contribute to *MLCSs* of $T$ are retained. To this end, all the points from $D^{k+1}_{init}$ are sorted by the best non-dominated sorting method in [23] to achieve its first frontier set, denoted as $(D^{k+1}_{init})_{1st}$. That is, $\forall p \in (D^{k+1}_{init})_{1st}$, there is no other point $p' \in D^{k+1}_{init}$ that dominates $p$. All points except $(D^{k+1}_{init})_{1st}$ are deleted from the set.

4) Through extensive experiments, we find that there are still many non-critical points residing in the $(D^{k+1}_{init})_{1st}$ although a large number of redundant points have been deleted in the above step. To eliminate the remaining redundant points, all the points in $(D^{k+1}_{init})_{1st}$ are further evaluated by Eq. 1. Since the points with the higher scores probably have little or no contribution to *MLCSs* to $T$, we only keep top $m$ points with the minimum values in $(D^{k+1}_{init})_{1st}$ and delete all the others. It is important to note that those deleted points may be key points, so this strategy leads to our algorithm being an approximate algorithm.

5) Let $k = k + 1$, and $D^k = (D^{k+1}_{init})_{1st}$, goto 2).

6) End the construction of *MLCS-ODAG*.

With the above steps, an optimized *MLCS-ODAG* of sequences $T$ with as few redundant points as possible are constructed with the forward iteration $D^k \rightarrow D^{k+1}$ procedure. An example of constructing *MLCS-ODAG* of 3 sequences is shown in Fig. 1.

$$Score(p) = \sum p_j/d + \max(p_j), (1 \leqslant j \leqslant d) \qquad (1)$$

where $\max(p_j)$ is the maximum value over all dimensions of $p$. The lower the value of $Score(p)$, the greater the likelihood that $p$ will contribute to *MLCSs* of the input sequences, and vice versa. The property of the proposed empirical function $Score(p)$ has been proved in [14, 19]. And it works well in our experiments.

**Mining all MLCSs.** Given the constructed *MLCS-ODAG*, we need to design an efficient and effective strategy to extract all *MLCSs* from it. We start by reviewing the following related concepts.

**Definition 4**: For a directed acyclic graph $G = \langle V, \preceq \rangle$, the *topological sorting* is to find an overall order of the vertices $V$ in $G$ from the partial order $\preceq$ [24].

**Definition 5**: A topological sorting algorithm [24] iteratively performs the following two steps until all vertices in $V$ have been traversed and processed: 1) outputting the vertices with in-degree 0; 2) deleting the edges connecting to the vertices.

Inspired by the topological sorting algorithm, after investigating our constructed *MLCS-ODAG*, we obtain the following important properties.

---

[3]The algorithms apply to the general case where the length of $S_i$ may not be the same. The only reason that we fix $|S_i| = n$ here is to facilitate the subsequent discussions.

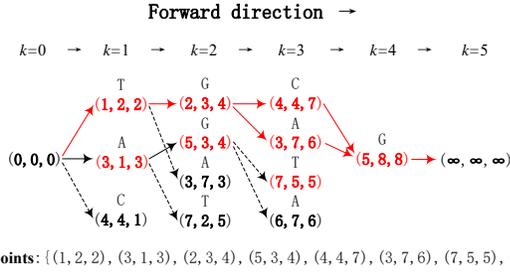

**key points:** $\{(1,2,2), (3,1,3), (2,3,4), (5,3,4), (4,4,7), (3,7,6), (7,5,5), (5,8,8)\}$

**Figure 1: *MLCS-ODAG* of** $S_1 = TGACGATC$, $S_2 = ATGCTCAG$ **and** $S_3 = CTAGTACG$ **over the alphabet** $\Sigma = \{A, C, G, T\}$, **in which the points are the points in** $(D_{init}^{k+1})_{1st}$ **of calculated results by the step 3) of the above constructing optimized *MLCS-DAG* procedure. The points to which dotted arrows point should be deleted after being evaluated by Eq. 1.**

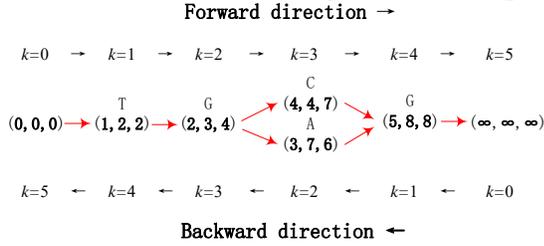

**Figure 2: The diagram of backward topological sorting to** *MLCS-ODAG* **by the proposed algorithm** *BackwardTopSort*.

**Theorem 1:** The sum of the numbers, denoted by $forward_l$ and $backward_l$ respectively, of the forward levels (*from source* $(0, 0, ..., 0)$ *to sink* $(\infty, \infty, ..., \infty)$) and the backward levels (*from the sink to the source*) of those points (*called key points*, denoted by $p^k$) *residing on the longest paths* corresponding to the *MLCSs* is exactly $|MLCS| + 1$. However, the critical points would not have the property (see Figs. 1 and 2). This can be formulated as follows:

$$forward_l(p^k) + backward_l(p^k) \equiv |MLCS| + 1 \quad (2)$$

**Proof:** Given a set of sequences, let their *MLCS* length in the *MLCS-ODAG* be $|MLCS|$, which is exactly equal to the maximum value of the forward levels in the *MLCS-ODAG* minus one (proven in [18, 25, 26]). Hence, given a key point $p^k$ residing on any of the longest paths of *MLCS-ODAG*, if its forward-level value is $x$, i.e., $forward_l(p^k) = x$, there must remain $|MLCS| - x$ levels from $p^k$ to the sink point. So, its backward-level value $backward_l(p^k)$ must be $|MLCS| - x + 1$. Hence, $forward_l(p^k) + backward_l(p^k) = x + (|MLCS| - x + 1) = |MLCS| + 1$. ∎

Based on the above observation, we replace the in-degree with the out-degree and layer the *MLCS-ODAG* by the topological sorting algorithm from the sink to the source, denoted as Algorithm *BackwardTopSort*. With this, all the non-critical points in *MLCS-ODAG* are now identified and can be easily removed. Fig. 2 shows the result of *BackwardTopSort* to Fig. 1. Note that the *MLCS-ODAG* shown in Fig. 2 contains only those key points, that is, each path in the *MLCS-ODAG* corresponds to an *MLCS* of $S_1$, $S_2$ and $S_3$. In addition, as shown in Fig. 2, some key points, such as point (7,5,5), would be deleted in the procedure of constructing *MLCS-ODAG*, resulting in some lost *MLCSs* of *MLCS-ODAG*, so our proposed *MLCS* algorithm, called *HA-MLCS* (High-precision Approximate MLCS Algorithm), belongs to the approximate *MLCS* algorithm category.

The pseudo-code of our algorithm *HA-MLCS* is given in Appendix D.

We compared our algorithm with the state-of-the-art algorithms *CRO* and *SA_MLCS* via extensive experiments. From the experimental results shown in Appendix B, we can draw the following conclusions: 1) Although the baseline *CRO* always has the fastest speed, it has the lowest precision; 2) Our algorithm *HA-MLCS* is much better than *SA_MLCS* in both running time and precision. In terms of running time, our algorithm is orders of magnitude faster than the baseline *SA_MLCS*; 3) Our *HA-MLCS* works well on big sequence data.

Notably, our *HA-MLCS* has the following unique properties:

**1) Low space-time complexity**

**Theorem 2:** The proposed algorithm *HA-MLCS* has $O(N \log N)$ time complexity and $O(dN + |E|)$ space complexity, respectively.

**Proof:** For each sequence $S_l$ of $T$ over the alphabet $\Sigma$ with length $n$, $O(n|\Sigma|)$ time is needed for constructing its successor table. The main operations in constructing the *MLCS-ODAG* consist mainly of the following. Firstly, establish the predecessor-successor relationships among dominants in *MLCS-ODAG*. Secondly, sort all the points in $D_{init}^{k+1}$ by Algorithm *BestNondominated-Sorting* [23]. Thus the time complexity for constructing the *MLCS-ODAG* is $O(N \log N)$, where $N = \sum_{k=1}^{|MLCS|-1} |D_{init}^{k+1}|$. The backward topological sorting on *MLCS-ODAG* by algorithm *Backward-TopSorting* needs $O(M)$, where $M$ is the total number of points in the final constructed *MLCS-ODAG*, and $M \ll N$. Therefore, the total time complexity of the proposed algorithm *HA-MLCS* is $O(dn|\Sigma|) + O(N \log N) + O(M) = O(N \log N)$ as $O(dn|\Sigma|) \ll O(N \log N)$. ∎

Similarly, the storage space of successor tables is $O(dn|\Sigma|)$, and the storage space of the *MLCS-ODAG* with $N$ points and $|E|$ edges is $O(dN + |E|)$. The space complexity of *HA-MLCS* is $O(dn|\Sigma| + dN + |E|) = O(dN + |E|)$ as $O(dn|\Sigma|) \ll O(dN + |E|)$.

**2) 100% *MLCS*' length precision**

**Theorem 3:** The precision of *MLCS* length is 100%, and the precision of the number of *MLCS* of *HA-MLCS* is calculated by Eq. 3.

$$P = 1 - C_{key} \sum_k (|(D_{init}^k)_{1st}| - m) / |K|, \quad (1 \leqslant k \leqslant |MLCS| - 1) \quad (3)$$

where, $C_{key}$ is the ratio of the total number of *key points* to the total number of points in *MLCS-ODAG*. $\sum_k (|(D_{init}^k)_{1st}| - m)$ denotes the set of points deleted in $\sum_k (D_{init}^k)_{1st}$ $(1 \leqslant k \leqslant |MLCS| - 1)$. The means of the notations $(D_{init}^k)_{1st}$ and $m$ are the same as before, shown in Sec. 2.3. $|K|$ represents the size of the set $K$ of the key points in *MLCS-ODAG*.

**Proof:** Since *the key points in MLCS-ODAG uniquely contribute to and determine both the length and the total number of mined MLCSs in MLCS-ODAG*, we argue that the precision of an approximate *MLCS* algorithm should be evaluated by both the mined *MLCS*' length and the number of *MLCS* of the algorithm. As the procedure for constructing *MLCS-ODAG* always keeps the frontier points of *MLCS-ODAG*, none of *MLCS*' length precision is lost, and the *MLCS*' length precision of *HA-MLCS* is 100%. However, since the deleted points $\sum_k (D_{init}^k)_{1st}$ may contain some key points, the number of mined *MLCS* precision of *HA-MLCS* is defined by Eq. 3. ∎

Notice that this property is very important for practical applications. In practice, it is not necessary to extract all the *MLCSs* between sequences, but to ensure that the length of the extracted *MLCSs* is accurate.

**3) A novel approximate *MLCS* algorithm suitable for big sequences analysis in practice**

The theoretical analysis and extensive experiments show that the proposed algorithm *HA-MLCS* is an efficient *MLCS* algorithm suitable for big sequences analysis in practice.

## 3 THE EVOLUTION OF COVID-19 VIRUS

### 3.1 COVID-19 sequences

We have collected nearly thirty thousand complete genome sequences of COVID-19 shown in Table 1 that are available publicly, covering 29,305 genomes isolated from COVID-19 in human hosts from 79 countries, 21 genomes from animals and the environment (outside the human bodies), 101 genomes from the four previously known flu-causing coronavirus, HCov-229E (3 genomes), HCov-OC43 (78 genomes), HCov-NL63 (16 genomes) and HCov-HKU1 (4 genomes), and 61 genomes from seven potentially lethal pathogenic viruses, SARS (11 genomes), MERS (11 genomes), Victoria (5 genomes), Lassa (6 genomes), Yamagata (5 genomes), Ebola (11 genomes), and Dengue (12 genomes). These sequences are downloaded from the following databases: GenBank or NCBI[4] (National Center for Biotechnology Information), GISAID[5] (Global Initiative on Sharing All Influenza Data), and CDC[6](Center for Disease Control and Prevention). The average sequence length is approximately 30,000.

### 3.2 Computing platforms and tools

Our investigation is carried out using two main computational tools, our proposed big sequence data analysis algorithm *HA-MLCS* (for similarity analysis) and existing MEGA X system[24] (for evolutionary relationship analysis). All the calculations were done on a computing cluster of 18 nodes (Intel(R) Xeon(R) Gold 5115 CPU, 2 chip, 10 cores/chip, 2 threads/core, @2.4 GHz and 96GB RAM).

### 3.3 Similarity metrics

Based on the similarity metric design criteria and a common method for extracting subsequences among sequences in bioinformatics and computational biology [19, 20], we give the following definitions and equations for computing the similarity of big sequences.

**Definition 6 (LD)**: Lowenstein/edit distance *LD* [22, 23, 24] is the minimum number of operands required to convert a character sequence $S_i$ to another sequence $S_j$ using the operations of inserting, deleting or changing a character. *LD* is the most commonly used measure of similarity [20] between two sequences, on which the similarity between a pair of sequences $S_i$ and $S_j$ is defined as

$$sim(S_i, S_j) = 1 - LD/min(|S_i|, |S_j|) \qquad (4)$$

The *LCS*-based similarity of a pair of sequences $S_i$ and $S_j$ is defined as

$$sim(S_i, S_j) = |LCS|/max(|S_i|, |S_j|) \qquad (5)$$

where |*LCS*| represents the length of the *LCSs* mined from the pair of sequences $S_i$ and $S_j$. $|S_i|$ and $|S_j|$ represent the lengths of sequences $S_i$ and $S_j$, respectively.

We use two similarity metrics/measures for each analysis experiment, one based on Lowenstein/edit distance *LD* (Eq. 4) and the other based on *LCS* (Eq. 5). We used the two similarity metrics to represent the similarities between a set of sequences, which can reveal some potential biological evolutionary or genetic relationships of different species quantitatively, enabling medical professionals and biological researchers to perform cross-verification or cross-comparison, and possibly deciding which method makes more biological sense.

### 3.4 Evolution and diversity of COVID-19

*3.4.1 Evolution of COVID-19 viruses from 79 countries.* In order to more accurately reveal the evolutionary relationship between the nearly 30,000 collected COVID-19 stains in 79 countries from December 2019 to May 2020, we first select all 25 sequences from China, the first country that reported the COVID-19 outbreak, from December 23, 2019 to December 31, 2019, and all the 401 sequences from China and other 18 countries in January 2020, which gives us a total of 426 earliest sequences out of all the collected nearly thirty thousand COVID-19 sequences from 79 countries. Then, we fed these 426 COVID-19 sequences into MEGA X to construct the evolutionary tree. From the constructed evolutionary tree, we selected all of the first generation sequences. After that, by the uniform random sampling method, i.e., by ensuring the sequences from each of 79 countries and their earliest sequences from Dec. 2019 to May 2020 can be drawn, we randomly sampled 10 groups of sequences from 79 countries between February and May, 2020, respectively. Then we added some new sequences with high confidence in each group to replace the low-confidence sequences with multiple *N* placeholders (*N* means the number of unknown characters). Finally, each group of sequences with all of the first generation's sequences calculated previously were fed into MEGA X to generate their evolution trees, respectively. 10 evolutionary trees produced by MEGA X demonstrated a high degree of consistency. Due to space limitations, we only present one evolutionary tree here in Fig.3 and Appendix C, and others are available at https://github.com/HA-MLCS/HA-MLCS/tree/master/supplementary_materials.

Investigating the evolutionary tree shown in Fig. 3 allows us to make the following observations:

1) Although China was the first country to report COVID-19 outbreaks and to provide COVID-19 sequences, none of the sequences were the earliest generations, and they were concentrated in the sixth and the eighth branches of the later generations in the tree.

2) Apart from the two sequences from Russia in the third branches and Spain in the forth branches of the tree, respectively, all the sequences from the top 15 countries currently reported to have the most severe outbreaks reside in the later generations in the fifth to tenth branches of the tree.

3) Of all the existing 29,305 COVID-19 sequences in human hosts from 79 countries, the earliest sequence No. GWHABKF00000000 2019,12.23 from Wuhan, China was sampled on Dec. 23, 2019. But it

**Table 1: The COVID-19 sequences in human hosts from 79 countries and other related viruses.**

| Country | 2019.12 | 2020.1 | 2020.2 | 2020.3 | 2020.4 | 2020.5 | Totality of sequences |
|---------|---------|--------|--------|--------|--------|--------|-----------------------|
| USA | 0 | 21 | 113 | 4271 | 1968 | 15 | 6388 |
| England | 0 | 2 | 37 | 6306 | 7624 | 276 | 14245 |
| Spain | 0 | 0 | 12 | 474 | 19 | 0 | 505 |
| Italy | 0 | 5 | 5 | 57 | 18 | 0 | 85 |
| France | 0 | 9 | 13 | 330 | 36 | 0 | 388 |
| Germany | 0 | 9 | 23 | 110 | 40 | 0 | 182 |
| India | 0 | 2 | 0 | 116 | 180 | 46 | 344 |
| Canada | 0 | 4 | 7 | 145 | 36 | 0 | 192 |
| China | 25 | 286 | 241 | 103 | 6 | 0 | 661 |

**Asia:** Bangladesh(20), Brunei(5), Cambodia(1), China(661), Georgia(15), India(344), Indonesia(9), Iran(6), Israel(223), Japan(130), Jordan(28), Kazakhstan(4), Kuwait(8), Lebanon(10), Malaysia(15), Nepal(1), Pakistan(3), Philippines(5), Qatar(16), Saudi Arabia(112), Singapore(157), South Korea(36), Sri Lanka(4), Thailand(118), Turkey(64), United Arab Emirates(25), Vietnam(19). **Countries:27; Sequences:2039**

**Europe:** Austria(237), Belarus(2), Belgium(488), Croatia(7), Czech(35), Denmark(584), England(14245), Estonia(5), Finland(41), France(388), Germany(182), Greece(135), Hungary(32), Iceland(505), Ireland(18), Italy(85), Latvia(25), Lithuania(3), Luxembourg(257), Netherlands(556), Norway(48), Poland(26), Portugal(100), Romania(2), Russia(207), Serbia(4), Slovakia(4), Slovenia(5), Spain(505), Sweden(163), Switzerland(74). **Countries: 31; Sequences: 18968**

**Africa:** Algeria(3), Congo(111), Egypt(2), Gambia(3), Ghana(15), Nigeria(1), Senegal(22), South Africa(19). **Countries: 8; Sequences: 176**

**North America:** Canada(192), Costa Rica(6), Jamaica(8), Mexico(17), Panama(1), USA(6388). **Countries: 6; Sequences: 6612**

**South America:** Argentina(28), Brazil(62), Chile(144), Colombia(83), Uruguay(9). **Countries: 5; Sequences: 326**

**Oceania:** Australia(1176), New Zealand(8). **Countries: 2; Sequences: 1184**

**Different hosts:** environment(2), rhinolophine(13), pangolin(6). Sequences: 21

**Other viruses with human as host:**
**HCov viruses:** HCov-229E(3), HCov-OC43(78), HCov-NL63(16), HCov-HKU1(4). **Sequences: 101**
**Seven pathogenic viruses:** SARS(11), MERS(11), Victoria(11), Lassa(12), Yamagata(11), Ebola(12), Dengue(12). **Sequences: 80**

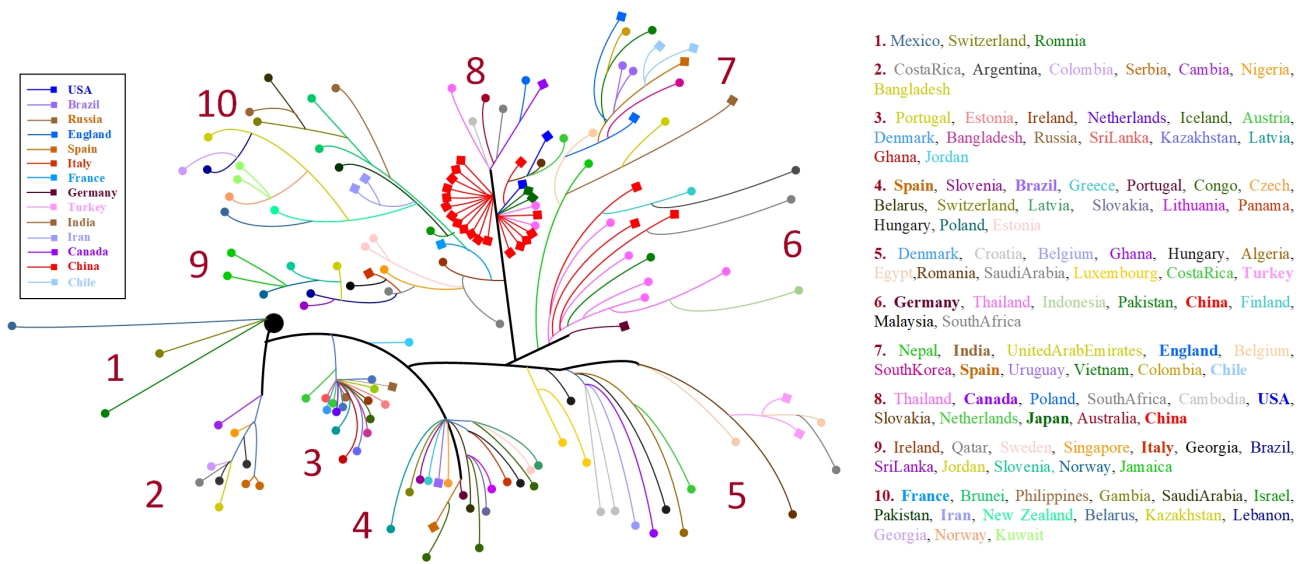

1. Mexico, Switzerland, Romnia

2. CostaRica, Argentina, Colombia, Serbia, Gambia, Nigeria, Bangladesh

3. Portugal, Estonia, Ireland, Netherlands, Iceland, Austria, Denmark, Bangladesh, Russia, SriLanka, Kazakhstan, Latvia, Ghana, Jordan

4. **Spain**, Slovenia, Brazil, Greece, Portugal, Congo, Czech, Belarus, Switzerland, Latvia, Slovakia, Lithuania, Panama, Hungary, Romania, Estonia

5. Denmark, Croatia, Belgium, Ghana, Hungary, Algeria, Egypt, Romania, SaudiArabia, Luxembourg, CostaRica, Turkey

6. **Germany**, Thailand, Indonesia, Pakistan, **China**, Finland, Malaysia, SouthAfrica

7. Nepal, India, UnitedArabEmirates, **England**, Belgium, SouthKorea, **Spain**, Uruguay, Vietnam, Colombia, Chile

8. Thailand, **Canada**, Poland, SouthAfrica, Cambodia, **USA**, Slovakia, Netherlands, **Japan**, Australia, **China**

9. Ireland, Qatar, Sweden, Singapore, **Italy**, Georgia, Brazil, SriLanka, Jordan, Slovenia, Norway, Jamaica

10. **France**, Brunei, Philippines, Gambia, SaudiArabia, Israel, Pakistan, Iran, New Zealand, Belarus, Kazakhstan, Lebanon, Georgia, Norway, Kuwait unexpectedly resided in the eighth branches of the later generations in the tree, which indicates that COVID-19 virus probably began to spread among people in multiple countries as early as December 2019. This is also confirmed in a recent study [11].

**Figure 3: The evolution tree of COVID-19 sequences in human hosts from different 79 countries, in which there are 10 branches denoted 1 to 10, respectively. In each branch, COVID-19 sequences from different countries shown on the right side of the figure are represented with different color dots. The 14 countries with the worst outbreaks in the legend on the left are indicated by square dots in different colors.**

4) The earliest sampled sequences from the 79 countries are distributed in different branches of the tree, which indicates the widespread infections and diversity of the COVID-19 virus in the world due to traveling and other reasons.

5) Although there is not yet enough evidence to trace COVID-19's origin, investigating the earliest generations' sequences in this evolution tree may provide some clues.

**Table 2: Average similarities (LCS-based/LD-based) between COVID-19 strains in human hosts of different months from China, Italy, USA and England, respectively.**

| China | | | | | | |
|---|---|---|---|---|---|---|
| Time | 2019.12 | 2020.01 | 2020.02 | 2020.03 | 2020.04 | 2020.05 |
| 2019.12 | 1 | 0.9967/0.9949 | 0.9947/0.9929 | 0.9920/0.9897 | 0.9960/0.9950 | – |
| 2020.01 | 0.9967/0.9949 | 1 | 0.99596/0.9939 | 0.9931/0.9904 | 0.9945/0.9936 | – |
| 2020.02 | 0.9947/0.9929 | 0.9959/0.9939 | 1 | 0.9946/0.9914 | 0.9928/0.9921 | – |
| 2020.03 | 0.9920/0.9897 | 0.9931/0.9904 | 0.9946/0.9914 | 1 | 0.9909/0.9904 | – |
| 2020.04 | 0.9962/0.9951 | 0.9945/0.9936 | 0.9928/0.9921 | 0.9909/0.9904 | 1 | – |
| 2020.05 | – | – | – | – | – | – |

| Italy | | | | | | |
|---|---|---|---|---|---|---|
| Time | 2019.12 | 2020.01 | 2020.02 | 2020.03 | 2020.04 | 2020.05 |
| 2019.12 | – | – | – | – | – | – |
| 2020.01 | – | – | 0.9984/0.9982 | 0.9969/0.9966 | 0.9976/0.9962 | – |
| 2020.02 | – | 0.9984/0.9982 | 1 | 0.9957/0.9954 | 0.9986/0.9978 | – |
| 2020.03 | – | 0.9969/0.9966 | 0.9957/0.9954 | 1 | 0.9954/0.9943 | – |
| 2020.04 | – | 0.9976/0.9962 | 0.9986/0.9978 | 0.9954/0.9943 | 1 | – |
| 2020.05 | – | – | – | – | – | – |

| USA | | | | | | |
|---|---|---|---|---|---|---|
| Time | 2019.12 | 2020.01 | 2020.02 | 2020.03 | 2020.04 | 2020.05 |
| 2019.12 | – | – | – | – | – | – |
| 2020.01 | – | 1 | 0.9920/0.9910 | 0.9879/0.9864 | 0.9932/0.9921 | 0.9959/0.9942 |
| 2020.02 | – | 0.9920/0.9910 | 1 | 0.99156/0.9868 | 0.9938/0.9890 | 0.9923/0.9892 |
| 2020.03 | – | 0.9879/0.9864 | 0.99156/0.9868 | 1 | 0.9909/0.9867 | 0.9888/0.9858 |
| 2020.04 | – | 0.9932/0.9921 | 0.9938/0.9890 | 0.9909/0.9867 | 1 | 0.9940/0.9913 |
| 2020.05 | – | 0.9959/0.9942 | 0.9923/0.9892 | 0.9888/0.9858 | 0.9940/0.9913 | 1 |

| England | | | | | | |
|---|---|---|---|---|---|---|
| Time | 2019.12 | 2020.01 | 2020.02 | 2020.03 | 2020.04 | 2020.05 |
| 2019.12 | – | – | – | – | – | – |
| 2020.01 | – | 1 | 0.9928/0.9927 | 0.9918/0.9917 | 0.9928/0.9927 | 0.9955/0.9941 |
| 2020.02 | – | 0.9928/0.9927 | 1 | 0.9958/0.9929 | 0.9951/0.9903 | 0.9935/0.9911 |
| 2020.03 | – | 0.9918/0.9917 | 0.9958/0.9929 | 1 | 0.9946/0.9902 | 0.9926/0.9905 |
| 2020.04 | – | 0.9928/0.9927 | 0.9951/0.9903 | 0.9946/0.9902 | 1 | 0.9935/0.9912 |
| 2020.05 | – | 0.9955/0.9941 | 0.9935/0.9911 | 0.9926/0.9905 | 0.9935/0.9912 | 1 |

Symbol '–' indicates that no sequence is provided.

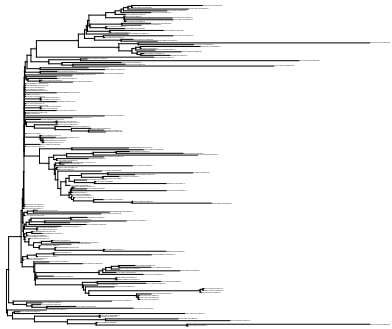

**Figure 4: The evolution tree of COVID-19 viruses from China.**

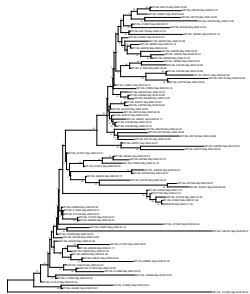

**Figure 5: The evolution tree of COVID-19 viruses from Italy.**

*3.4.2 Similarity and evolution of COVID-19 viruses.* In this study, we calculated all similarities of COVID-19 viruses among themselves and also between COVID-19 viruses and other related viruses. Notice that the similarity matrix of homogeneous sequences is a symmetric matrix, which represents pairwise comparisons of two sequences of the same virus type; otherwise an asymmetric matrix, which represents pairwise comparisons of two sequences of two different virus types. Note that each sequence in the symmetric/asymmetric matrix represents a common sequence of 5 sequences calculated by the proposed algorithm *HA-MLCS*. The average similarity between sequences in the same virus type is computed using all the elements of the upper/lower half of the symmetric similarity matrix except the diagonal elements, while the average similarity between sequences of two different virus classes is calculated using all the elements of the asymmetric similarity matrix.

Since China was the first country that reported the COVID-19 outbreak and submitted COVID-19 viruses, and USA, Italy and England are the countries most affected by the COVID-19 pandemic with a lot of sequences from Jan. to May 2020, the similarity and evolutionary analysis of the sequences of the above four countries are specially reported here[7], which are shown in Table 2, Figs. 4-5 and Appendix C, respectively. From the above analysis, we can make the following observations:

1) Although the overall similarities of these human strains are high, we observed a reduction of the similarities in later months of all the above four countries, indicating mutations within the human population is already occurring.

2) Averages of nucleotide differences from the four countries are 286.39, 292.35, 268.49 and 247.61, respectively, corresponding to averages of nucleic acid mutation points 325, 423, 378 and 289, respectively. These changes imply rapid evaluations of this virus, which might result in attenuation or more virulent strains. All these differences are statistically significant ($p < 0.0013$), which indicates that COVID-19 has begun its divergence in the human population.

3) Although the sequences of COVID-19 virus from the above four countries have evolved at different rates, all the different countries' viruses are steadily mutating, which potentially explains the underlying differences in virulence and alerts us to consider this divergence in designing antibodies and vaccines.

4) By investigating the sequence locations in their evolutionary tree of the above countries, as well as all other countries, we can infer that the first generation sequences is positively related to their sampling time, but not entirely. In addition, for each country, there are also some outlier sequences, e.g., strain EPI ISL 417180 China 2020.02.03. Further research on the first generation strains including outliers from these countries will have important significance in searching for the virus transmission path.

*3.4.3 The similarity between COVID-19 and related viruses.* To help trace the original or the intermediate host of COVID-19 and to assist the finding of natural remedies, we analyzed the similarity between COVID-19 viruses in different hosts, including human, rhinolophine, pangolin, and environmentally collected strains.

We found that COVID-19 virus living in the environment is highly similar to that living in the human body. The average similarity can reach 0.9972/0.9972 (*LCS/LD*). This is expected as this is likely to reflect what is being transmitted right now among the human population. We also found strong similarities between TG13 and RaTG13 (rhinolophine host) and COVID-19 (human host), reaching 0.9599/0.9584 (*LCS/LD*) and 0.9599/0.9585 (*LCS/LD*), respectively. But the average similarities between COVID-19 with human host and the other COVID-19 strains with rhinolophine host are not very high, 0.7416/0.6631, lower than the similarity with COVID-19 strains (pangolin host), 0.8742/0.8604, by 13% and 20% (*LCS/LD*).

---

[7] The sequence similarity matrices and evolutionary analysis of other heavily affected countries are available at https://github.com/HA-MLCS/HA-MLCS.

It has been reported that many symptoms of COVID-19 patients resemble those of the influenza patients infected with one of the four known flu coronaviruses. Therefore, we computed the sequence similarities between COVID-19 viruses and the four known flu coronaviruses, HCov-229E, HCov-OC43, HCov-NL63 and HCov-HKU1. The average similarity matrices, computed with Eq. 4 (*LD-based*) and Eq. 5 (*LCS-based*), respectively, are shown in Appendix E. The average similarities, are 0.6532/0.5557 for HCov-229E, 0.6806/0.5619 for HCov-OC43, 0.6597/0.5607 for HCov-NL63 and 0.6909/0.5612 for HCov-HKU1. We observed that the difference in the similarity values between the two (*LCS* and *LD*) metrics is about 10%, but the trends of the two results are consistent.

Compared to the shared similarity between COVID-19 and the seven lethal strains, the similarity between COVID-19 and other known flu-causing coronaviruses is in general higher except SARS and MERS. This pinpoints the importance of revisiting the treatment of flus and studying whether drug repurposing could possibly alleviate the current COVID-19 crisis.

It is also worth noting that the similarities between COVID-19 strains and viruses HCov-OC43, Lassa, MERS, Victoria, Yamagate, Ebola and Dengue have increased steadily over the past six months.

## 4 CONCLUSION

Pathogenic mechanism, virus detection, and vaccine and drug developments all heavily depend on the analysis of the complete genome sequences of COVID-19. This study provides important information to support the decision making of medical and healthcare professionals in tracking COVID-19's mutation paths, developing virus detection tools, vaccines and drugs, and controlling the epidemic. Below, we reiterate several key findings.

First, the genome sequences of COVID-19 viruses in humans have already gone through mutations over the past six months. This has important implication for developing COVID-19 test kits, vaccines and antibody treatments. Recently, efforts to isolate antibodies for COVID-treatment have been announced by several pharmaceutical companies, and vaccines are being actively developed by many research labs around the world. The breadth of the coverage of the antibodies and vaccines will be critical in determining its efficacy.

Second, COVID-19 shares little similarity with Ebola, but more with the four previously known flu-causing coronaviruses (HCov-229E, HCov-OC43, HCov-NL63 and HCov-HKU1), and even more with SARS. The sequence analysis suggests that treatments to SARS and other flu-inducing coronaviruses might be another roadmap that we should explore. We recommend considering this during medication and treatment development.

Third, COVID-19 virus strains from most countries might have gone through multiple evolution paths. Extensive analyses of COVID-19 strains from different countries potentially lead us to find the first generation COVID-19 virus and its origin. As the data shown here, at the national scale, COVID-19 could have already spread through multiple routes. This also highlights the need to develop more aggressive isolation and quarantine procedures for anyone demonstrating suspicious symptoms, even without direct or known contact with a patient.

## ACKNOWLEDGMENTS

The work of Y. Li, J. Cui, Y. Shen, Y. Xu and X. Ma was partly supported by the NSFC (No. 61472296, 61976168, 61702391, and 61772394) and the NKRDPC (No. 2018YFE0207600).

# APPENDIX

## A   MAIN DATA STRUCTURE

**Successor tables (ST).** The successor tables $\{ST_1, ST_2, ..., ST_d\}$ of the sequence set $T = \{S_1, S_2, ..., S_d\}$ are built to support the compression of the data and quick search for the immediate successors of the points. For a sequence $S_l = x_1, x_2, ..., x_n$ from the sequence set $T$ over a finite alphabet $\Sigma = (c_1, c_2, ..., c_k)$, its successor table $ST_l$ is a two-dimensional array, where $ST_l[i, j]$ (the element of the $i$th row and the $j$th column) is defined as

$$ST_l[i, j] = \begin{cases} min\{r|x_r = c_i, r \geqslant 1, r \geqslant j, 1 \leqslant i \leqslant |\Sigma|, \\ 0 \leqslant j \leqslant n\} \end{cases} \quad (6)$$

From Eq. 6, we can see that $ST_l[i, j]$ denotes the minimal position $r$ (the $r$th character position) of the sequence $S_l$ with $x_r = c_i$ after position $j$. See the examples in Fig. 6.

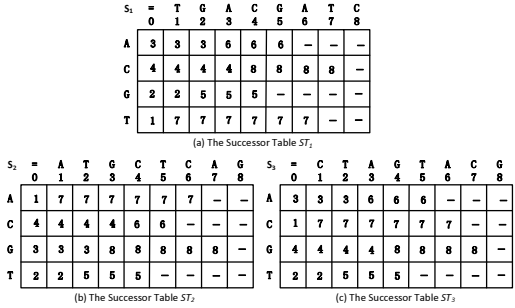

**Figure 6: The constructed successor tables $ST_1, ST_2$ and $ST_3$ corresponding to the sequences $S_1, S_2$ and $S_3$ (in the paper), where "−" indicates ∅.**

The set $S_{suc}$ of immediate successors of a $d$-dimensional point $p = (p_1, p_2, ..., p_d)$ can be obtained efficiently in $O(d|\Sigma|)$ time. For a $d$-dimensional point $p$, the operation for producing its $S_{suc}$ can be characterized by Eq. 7.

$$S_{suc} = \{(ST_1[i', p_1], ST_2[i', p_2], ..., ST_d[i', p_d])\}$$
$$s.t. 1 \leqslant i' \leqslant |\Sigma|, \forall ST_l[i', p_i] \neq \emptyset, 1 \leqslant l, i \leqslant d \quad (7)$$

For example, for the dominant $(2, 3, 4)$ of the sequences $S_1, S_2$ and $S_3$ (in the paper), we can couple the corresponding rows 1-4 of the second, third and forth columns from the successor tables $ST_1, ST_2$ and $ST_3$ to obtain all its immediate successors $(3, 7, 6)$, $(4, 4, 7)$, $(5, 8, 8)$ and $(7, 5, 5)$ corresponding to the characters A, C, G, and T, respectively. There is no immediate successor for the dominant $(6, 7, 3)$ due to the coupling results $(\_, \_, 6)$, $(8, \_, 7)$, $(\_, 8, 4)$ and $(7, \_, 5)$, which indicates none of the points is an immediate successor according to Eq. 7.

## B   EXPERIMENTAL RESULTS

We evaluate the performance of the approximate algorithms, whose performances vary in terms of not only efficiency but also precision. Here, the precision is measured by Eq. 3. The results for all the tested approximate algorithms, including the state-of-the-art CRO, SA_MLCS as well as our algorithm HA-MLCS are shown in Tables 3 and 4.

**Table 3: Precisions ($P$) and running times ($T$) of *CRO* (A1), *SA-MLCS* (A2) and *HA-MLCS* (A3) for 5 sequences with various lengths ($|S_l|$).**

| | | | $|\Sigma|$=4 | | | | | | | $|\Sigma|$=20 | | | |
|---|---|---|---|---|---|---|---|---|---|---|---|---|---|
| | A1 | | A2 | | A3 ($m$ = 100) | | | A1 | | A2 | | A3 ($m$ = 100) | |
| $|S_l|$ | $P$ | T(s) | $P$ | T(s) | $P$ | T(s) | $|S_l|$ | $P$ | T(s) | $P$ | T(s) | $P$ | T(s) |
| 1.0E+3 | 0.524 | 0.08 | 0.822 | 0.97 | 0.988 | 0.71 | 1.0E+3 | 0.411 | 0.09 | 0.893 | 7.43 | 0.992 | 0.63 |
| 2.0E+3 | 0.496 | 0.11 | 0.818 | 1.94 | 0.981 | 1.42 | 2.0E+3 | 0.366 | 0.12 | 0.877 | 17.31 | 0.981 | 1.26 |
| 5.0E+3 | 0.391 | 1.01 | 0.792 | 2.41 | 0.985 | 1.35 | 5.0E+3 | 0.357 | 5.79 | 0.857 | 47.99 | 0.978 | 2.90 |
| 1.0E+4 | 0.347 | 3.48 | 0.743 | 9.33 | 0.975 | 5.50 | 1.0E+4 | 0.334 | 2.27 | 0.822 | 113.70 | 0.967 | 4.35 |
| 2.0E+4 | 0.322 | 5.86 | 0.717 | 32.59 | 0.972 | 9.22 | 2.0E+4 | 0.327 | 8.67 | 0.802 | 239.00 | 0.971 | 8.12 |
| 5.0E+4 | 0.314 | 33.01 | 0.701 | 87.06 | 0.968 | 27.56 | 5.0E+4 | 0.305 | 53.94 | 0.791 | 541.80 | 0.965 | 18.79 |
| 1.0E+5 | 0.296 | 133.20 | 0.678 | 156.79 | 0.959 | 52.86 | 1.0E+5 | 0.279 | 218.00 | 0.751 | 5272.00 | 0.952 | 39.87 |
| 1.0E+7 | + | + | 0.269 | 1609.00 | 0.951 | 557.20 | 1.0E+7 | + | + | 0.741 | 44190.00 | 0.962 | 403.60 |
| 1.0E+8 | + | + | + | + | 0.948 | 3430.00 | 1.0E+8 | + | + | + | + | 0.963 | 2985.00 |

Symbol '+' indicates the memory overflow leading to calculating failure.

**Table 4: Precisions ($P$) and running times ($T$) of *CRO* (A1), *SA-MLCS* (A2) and *HA-MLCS* (A3) for $d$ sequences with lengths ($|S_l|$) 1000 and 2000, respectively.**

| | | | $|\Sigma|$=4, $|S_l|$=1000 | | | | | | | $|\Sigma|$=20, $|S_l|$=2000 | | | |
|---|---|---|---|---|---|---|---|---|---|---|---|---|---|
| | A1 | | A2 | | A3 ($m$ = 100) | | | A1 | | A2 | | A3 ($m$ = 100) | |
| $d$ | $P$ | T(s) | $P$ | T(s) | $P$ | T(s) | $d$ | $P$ | T(s) | $P$ | T(s) | $P$ | T(s) |
| 10 | 0.689 | 0.03 | 0.792 | 0.62 | 0.988 | 0.41 | 10 | 0.698 | 0.03 | 0.801 | 0.12 | 0.987 | 0.08 |
| 50 | 0.666 | 0.06 | 0.756 | 0.82 | 0.977 | 0.51 | 50 | 0.671 | 0.06 | 0.811 | 0.11 | 0.971 | 0.20 |
| 100 | 0.645 | 0.08 | 0.758 | 0.97 | 0.975 | 0.63 | 100 | 0.655 | 0.07 | 0.801 | 0.19 | 0.978 | 0.13 |
| 400 | 0.632 | 0.17 | 0.742 | 1.02 | 0.973 | 0.74 | 400 | 0.589 | 0.17 | 0.812 | 0.23 | 0.975 | 0.19 |
| 800 | 0.617 | 0.19 | 0.738 | 1.33 | 0.973 | 0.88 | 1000 | 0.623 | 0.33 | 0.805 | 0.32 | 0.972 | 0.25 |
| 1000 | 0.602 | 0.26 | 0.717 | 2.26 | 0.969 | 1.24 | 5000 | 0.611 | 1.51 | 0.798 | 0.41 | 0.966 | 0.35 |
| 3000 | 0.583 | 0.45 | 0.703 | 3.11 | 0.967 | 1.64 | 10000 | 0.594 | 2.98 | 0.785 | 0.43 | 0.966 | 0.41 |
| 5000 | 0.577 | 0.70 | 0.695 | 3.57 | 0.965 | 1.76 | 14000 | 0.568 | 4.26 | 0.773 | 0.50 | 0.964 | 0.47 |
| 10000 | 0.534 | 1.11 | 0.686 | 3.83 | 0.959 | 2.18 | 20000 | 0.536 | 5.57 | 0.758 | 1.21 | 0.959 | 0.73 |

## C   THE EVOLUTION TREES OF COVID-19 VIRUSES FROM THE USA AND ENGLAND

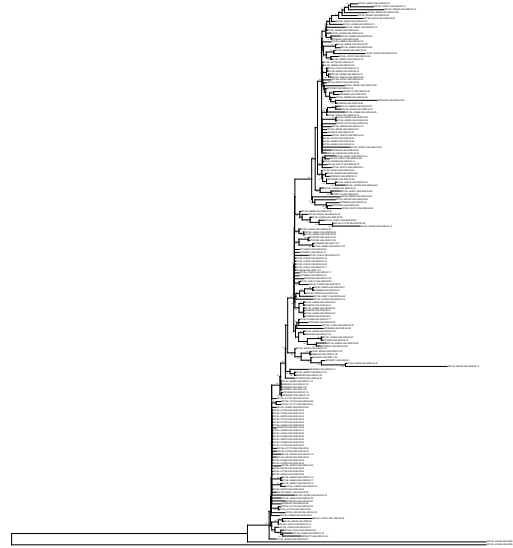

**Figure 7: The evolution tree of COVID-19 viruses in human hosts from USA.**

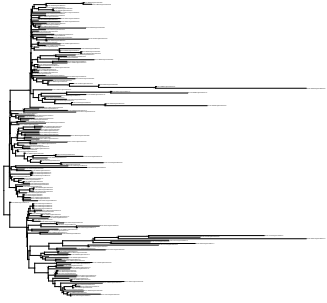

**Figure 8: The evolution tree of COVID-19 viruses in human hosts from England.**

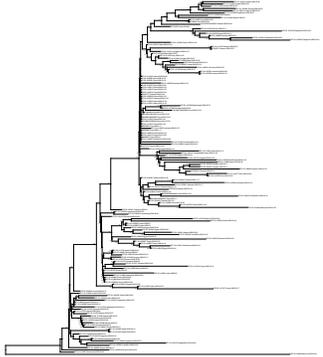

**Figure 9: The evolution tree of COVID-19 viruses in human hosts from different 79 countries.**

## D ALGORITHM PSEUDOCODE

---

**Algorithm 1** HA-MLCS($T, \Sigma$)

---

1: $ST \leftarrow$ Construct Successor Tables of sequence set $T$ with $\Sigma$;
2: $k \leftarrow 0$; $MLCS\text{-}ODAG \leftarrow \varnothing$; $D^k \leftarrow \{(0, 0, ..., 0)\}$;
3: $D_{init}^{k+1} \leftarrow Successor(D^k, ST)$; //calculate successor point set $D_{init}^{k+1}$ of $D^k$
4: **while** $D_{init}^{k+1} \neq \varnothing$ **do** //Constructing $MLCS\text{-}ODAG$ with layer by layer
5:    $(D_{init}^{k+1})_{1st} \leftarrow$ BestNondominatedSorting($D_{init}^{k+1}$);
6:    Calculate score($p$) by Eq. 1, $p \in (D_{init}^{k+1})_{1st}$;
7:    $D_{1st}^{k+1} \leftarrow$ Keep top $m$ points with the minimum score in $(D_{init}^{k+1})_{1st}$;
8:    $MLCS\text{-}ODAG \leftarrow MLCS\text{-}ODAG \cup D_{1st}^{k+1}$; $k \leftarrow k + 1$;
9:    $D_{init}^{k+1} \leftarrow Successor(D^k, ST)$;
10: **end while**
11: $maxlevel \leftarrow k - 1$; $k \leftarrow 0$; $D^0 \leftarrow \{(\infty, \infty, ..., \infty)\}$;
12: **while** $D^k \neq \varnothing$ **do** //Algorithm BackwardTopSorting
13:    $D^{k+1} \leftarrow \varnothing$;
14:    **for** $q \in D^k$ **do**
15:      **for** $p \in precursor[q]$ **do**
16:        **if** $tlevel[p] + k \neq maxlevel$ **then**
17:          Delete $p$ from $MLCS\text{-}ODAG$;
18:        **else**
19:          $D^{k+1} \leftarrow D^{k+1} \cup \{p\}$;
20:        **end if**
21:      **end for**
22:    **end for**
23:    $k \leftarrow k + 1$;
24: **end while**
25: **return** all of the $MLCSs$ of sequence set $T$;

---

The proposed algorithm *HA-MLCS* is implemented in Java JDK1.8. Where $m$ is a user-customized parameter ($1 \leqslant m \in \mathbb{Z}$), which represents how many number of key points to be retained in each layer when constructing *MLCS-ODAG*. The source code of algorithm *HA-MLCS* is available at: https://github.com/HA-MLCS/HA-MLCS

## E THE SIMILARITY BETWEEN COVID-19 AND RELATED VIRUSES

**Table 5: The similarity matrices.**

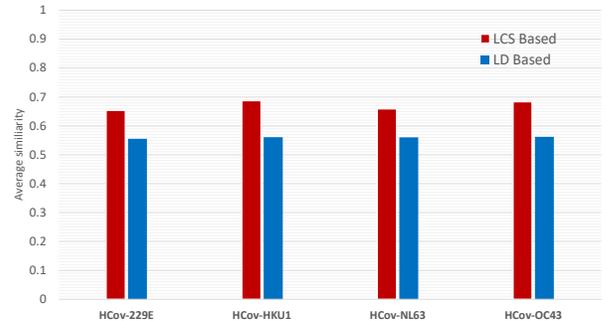

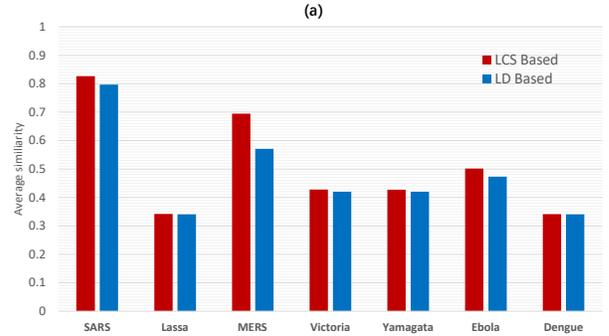

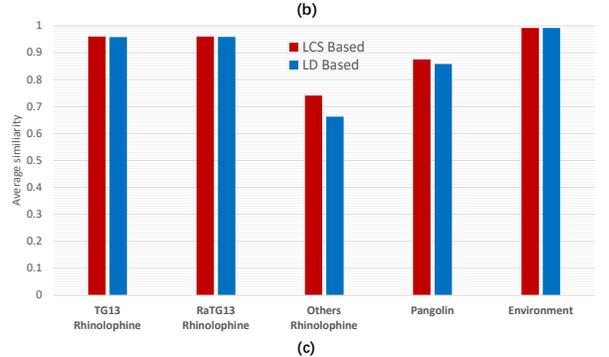

**Figure 10: The schematic diagram between COVID-19 strains and other viruses.**